\documentclass{aastex}
\usepackage{emulateapj5}

\newcommand{\oii}{[\ion{O}{2}] $\lambda$3727}
\newcommand{\oiiia}{[\ion{O}{3}] $\lambda$4959}
\newcommand{\oiiib}{[\ion{O}{3}] $\lambda$5007}
\newcommand{\zp}{\mbox{$z_{\mbox{\tiny phot}}$}}
\newcommand{\zs}{\mbox{$z_{\mbox{\tiny spec}}$}}
\newcommand{\sig}{\mbox{$\sigma_{\mbox{\tiny $\parallel$}}$}}
\newcommand{\eg}{e.g.}
\newcommand{\hr}{\mbox{$^h$}}
\renewcommand{\min}{\mbox{$^m$}}
\renewcommand{\deg}{\mbox{$^{\circ}$}}
\renewcommand{\arcmin}{\mbox{$^\prime$}}
\newcommand{\secper}{$\stackrel{\prime \prime}{\mbox{.}}$}
\renewcommand{\farcs}{$\stackrel{\prime \prime}{\mbox{.}}$}
\newcommand{\simgtr}{\raisebox{-.2ex}{$\stackrel{>}{\mbox{\tiny $\sim$}}$}}

\newcommand{\fsz}{\footnotesize}

\slugcomment{\it Accepted for Publication in the Astronomical Journal.}
\lefthead{Dawson et al.}
\righthead{Serendipitous Galaxies in the HDF}

\begin{document}

\title{Serendipitously Detected Galaxies in the Hubble Deep Field\altaffilmark{1}}

\author{
Steve Dawson\altaffilmark{2}, 
Daniel Stern\altaffilmark{3,4}, 
Andrew J. Bunker\altaffilmark{3,5}, 
Hyron Spinrad\altaffilmark{2}, 
and Arjun Dey\altaffilmark{6}}

\altaffiltext{1}{ 
Based on observations made at the W.M. Keck Observatory, which is operated
as a scientific partnership among the California Institute of Technology,
the University of California and the National Aeronautics and Space
Administration.  The Observatory was made possible by the generous
financial support of the W.M. Keck Foundation.}

\altaffiltext{2}{
Astronomy Department, University of California at Berkeley, CA 94720 USA,
email: (sdawson, spinrad)@astro.berkeley.edu}

\altaffiltext{3}{
formerly, Astronomy Department, University of California at Berkeley, CA 94720 USA}

\altaffiltext{4}{
Jet Propulsion Laboratory, California Institute of Technology, Mail Stop
169--327,
Pasadena, CA 91109 USA, email: stern@zwolfkinder.jpl.nasa.gov}

\altaffiltext{5}{
University of Cambridge, Institute of Astronomy, Madingley
Road, Cambridge CB3 0HA, UK, email: bunker@ast.cam.ac.uk}

\altaffiltext{6}{
KPNO/NOAO, 850 N. Cherry Ave., P.O. Box 26732, Tucson, AZ 85726 USA,
email: dey@noao.edu}


\begin{abstract}
We present a catalog of 74 galaxies detected serendipitously during a
campaign of spectroscopic observations of the Hubble Deep Field North
(HDF) and its environs.  Among the identified objects are five candidate  
Ly$\alpha$--emitters at $z \simgtr 5$, a galaxy cluster at $z = 0.85$, and
a Chandra source with a heretofore undetermined redshift of $z = 2.011$.
We report redshifts for 25 galaxies in the central HDF, 13 of which had no
prior published spectroscopic redshift.  Of the remaining 49 galaxies, 30
are located in the single--orbit HDF Flanking Fields.  We discuss the  
redshift distribution of the serendipitous sample, which contains galaxies
in the range $0.10 < z < 5.77$ with a median redshift of $z = 0.85$, and
we present strong evidence for redshift clustering.  By comparing our
spectroscopic redshifts to optical/IR photometric studies of the HDF, we
find that photometric redshifts are in most cases capable of producing
reasonable predictions of galaxy redshifts.  Finally, we estimate the   
line--of--sight velocity dispersion and the corresponding mass and
expected X--ray luminosity of the galaxy cluster, we present strong  
arguments for interpreting the Chandra source as an obscured AGN, and we
discuss in detail the spectrum of one of the candidate $z \simgtr 5$
Ly$\alpha$--emitters.
\end{abstract}



\section{Introduction}

The Hubble Deep Field North \citep[][hereafter W96]{williams96} ranks
among the most thoroughly studied portions of the extragalactic universe. 
The extremely deep multi--color images obtained with the WFPC2 camera on
the {\it Hubble Space Telescope}, reaching AB mag $B_{450} \sim 29$ with
0\farcs1 resolution, have revolutionized our understanding of the faint
galaxy population and have yielded diverse new results in observational
cosmology.  Follow--up observations to the original survey span the
electromagnetic spectrum, from the radio \citep{fomalont97, richards98,
richards00} to the sub--millimeter \citep{hughes98,barger00}, to both
ground and space--based near infrared \citep{hogg97, dickinson99,
dickinson00, thompson99} and far infrared \citep{aussel99}.  Recently,
X--ray data have become available \citep{hornschemeier00,hornschemeier01}, and UV
observations with the Space Telescope Imaging Spectrograph are in progress
\citep{gardner01}.  In addition to imaging, numerous groups are pursuing
spectroscopic observations of galaxies in the HDF.  \citet{cohen00} report
on a magnitude--limited sample more than 92\% complete to Vega mag $R =
24$;  \citet{steidel96a} and \citet{lowenthal97} report on color--selected
samples of Lyman--break galaxies at $z \sim 3$; while \citet{zepf_b97}
report on a morphologically selected sample of probable gravitational
lenses.  See \citet{ferguson00} for a review of measurements and
phenomenology of sources in the HDF across the electromagnetic spectrum. 

Consequently, the HDF and the eight adjacent, single--orbit $I_{814}$
Flanking Fields (see W96, Table 2) now constitute a very rich database for
the study of galaxy formation and evolution.  Early results included the
confirmation of a flattening in the slope of the faint elliptical/S0
galaxy number count--magnitude relation \citep{abraham96, zepf97}, as well
as the revealed inadequacy of the Hubble sequence as a classification
scheme for galaxies fainter than $I_{AB} > 24$ mag \citep{abraham96}.  The
selection of four very broad bandpass filters for the WFPC2 observations
was driven partly by the desire to identify high--redshift galaxies via
the Lyman--break technique.  Indeed, this strategy facilitated the
discovery of distant galaxies whose Lyman--breaks have been redshifted
into the $U$--band \citep{steidel96a, lowenthal97}, the $B$--band
\citep{steidel99}, and beyond \citep{spinrad98,weymann98}.  The exquisite
resolution of the WFPC2 images spurred considerable effort toward
quantifying galaxy morphology, leading to the disentanglement of
morphological $k$--correction from morphological evolution, and revealing
an increase in the fraction of true irregulars at faint magnitudes/high
redshift \citep{bunker99}.  Most recently, mining of this data--rich field
has yielded refined techniques in estimating photometric redshifts
\citep[\eg][]{fernandez99} and has produced dramatic implications for the
history of star--formation \citep{madau96, steidel99} as well as for the
role of dust in the distant universe \citep{hughes98, ouchi99}. 

We are pursuing a variety of programs to study distant galaxies in the
HDF.  The primary science from these observations, discussed elsewhere,
includes extremely deep ($\sim 10^{\rm h}$)  moderate and high--resolution
Keck/LRIS spectroscopy of Lyman--break galaxies at $z \sim 3$ aimed at
understanding their stellar populations and galactic dynamics
\citep[\eg][]{bunker98}, and low--resolution spectroscopy of $B$--band and
$V$--band dropouts whose colors suggest a population of galaxies with
Lyman--breaks and significant Ly$\alpha$--forest absorption at $z > 4$
\citep{spinrad98}.  In the course of these observations, we have targeted
more than 65 galaxies in the HDF and its environs for deep spectroscopy,
and in so doing we have serendipitously observed some 125 objects which
were located propitiously along the slit of a target.  Out of the sample
of serendipitous detections, we have determined redshifts for 74 sources,
with 25 galaxies in the HDF proper, 30 galaxies in the HDF Flanking
Fields, and 19 galaxies beyond but in the vicinity of the Flanking Fields. 
Thirteen of the detections in the central HDF provided the first ever
spectroscopic redshift determinations for those sources.

From the first detection of pulsars to the discovery of the cosmic
microwave background, serendipity has historically made significant
contributions to astronomy.  In extra--galactic astronomy in particular,
dramatic serendipitous detections include the discovery of a galaxy
cluster at $z=2.40$ \citep{pascarelle96}, at least three quasars at $z >
4$ \citep{mccarthy88, schneider94, schneider00}, and the discovery of the
first object at $z > 5$ \citep{dey98}.  Serendipity plays a less dramatic
but still significant role in large scale redshift surveys:  serendipitous
detections make up roughly 8\% of the measured galaxies in the complete
$K_s < 20$ mag galaxy sample presented by \citet{cohen99}.  Serendipitous
surveys in their own right are efficient, as they require no direct
initial allocation of telescope time, and they have proven to be both
competitive with and complementary to narrow--band imaging surveys.  See
\citet{thompson95}, \citet{manning00}, and \citet{stern00a} for reports on
serendipitous searches for high--redshift Ly$\alpha$ emission. 

Though none of the serendipitous detections reported herein constitute
singularly momentous discoveries, given the status of the HDF as ranking
among the most thoroughly mapped pieces of the extragalactic universe, we
would be remiss not to report all galaxy redshifts determined in the
course of our observations of this well--studied field.  In \S \ref{data}
we discuss the spectroscopic observations and the data reduction.  In \S
\ref{ids} we describe the redshift determination and the process by which
the serendipitously detected galaxies were visually identified.  We  
present the catalog of serendipitously detected galaxies in \S
\ref{catalogues}, and we discuss their distribution in redshift space, the
comparison between spectroscopic and photometric redshifts, the observed  
properties of the galaxy cluster at $z=0.85$, the observed properties of
the Chandra source at $z=2.011$, and the candidate $z \simgtr 5$
Ly$\alpha$--emitters in \S \ref{discuss}.  Throughout this paper, we adopt
an Einstein--de Sitter cosmology with $H_0 = 100 \, h_{100}$ km s$^{-1}$
Mpc$^{-1}$, $q_0 = 0.5$, and $\Lambda = 0$.  All quoted magnitudes are in
the AB system\footnote{ The AB magnitude system is defined such that
$m(\mbox{AB}) = -2.5 \log(f_{\nu}) - 48.60$ with $f_\nu$ measured in erg
s$^{-1}$ cm$^{-2}$ Hz$^{-1}$ \citep{oke1983}.  The value of the constant  
is set by the condition $m(\mbox{AB}) = V$ for a flat--spectrum source.}
unless otherwise specified.

\section{Observations and Data Reductions}
\label{data}

Between 1997 February and 2001 February, we obtained deep spectra of
photometrically selected high--redshift candidates in the HDF and its
environs.  The data were taken with the Low Resolution Imaging
Spectrometer (LRIS; Oke et al.\ 1995)  at the Cassegrain focus on the 10m
Keck I and Keck II telescopes.  The camera uses a Tek 2048$^2$ CCD
detector with a pixel scale of 0\farcs212 pixel$^{-1}$.  To maximize
observing efficiency, we exclusively used the dual amplifier readout mode.
Except for three longslit observations in 1997 February, the data were
taken with slitmasks designed to obtain spectra for $\sim 15$ targets
simultaneously.

For the vast majority of observations, we used a 150 lines mm$^{-1}$
grating blazed at 7500 \AA, which produces a $4.8$ \AA\, pix$^{-1}$
dispersion.  The spectral coverage with this grating is approximately 4000
\AA\, to 1 micron, allowing us observe the entire optical window
irrespective of the grating tilt or the location of the slit on the
slitmask.  We used a 300 lines mm$^{-1}$ grating blazed at 5000 \AA\ (2.55
\AA\, pix$^{-1}$ dispersion) on one set of observations, a 400 lines
mm$^{-1}$ grating blazed at 8500 \AA\ (1.86 \AA\, pix$^{-1}$ dispersion)
on two sets of observations, and a 600 lines mm$^{-1}$ grating blazed at
5000 \AA\ (1.28 \AA\, pix$^{-1}$ dispersion) on one set of observations.
For targets within the central HDF (where the astrometric solutions are
well--determined), we employed 1\farcs0 slits, yielding a spectral
resolution of $\lambda / \Delta \lambda_{\mbox{\tiny FWHM}} = 375$ with
the 150 lines mm$^{-1}$ grating.  For targets outside of the HDF (where   
the astrometric solutions are less well--determined), we employed 1\farcs5
slits, yielding a spectral resolution of $\lambda / \Delta
\lambda_{\mbox{\tiny FWHM}} = 250$ with the 150 lines mm$^{-1}$ grating.
The minimum set of exposures for any given target was 3 $\times$ 1800s.
As the position angle of the slit for a particular target normally changed
from observation to observation, only a small number ($\sim 5$)  of the
serendipitous detections benefited from re--observation.

The most recent set of observations (2001 February) were made with the
advent of the LRIS--B spectrograph channel \citep{mccarthy98}.  For these
observations, we used the 400 lines mm$^{-1}$ grating blazed at 8500 \AA\
in red channel, and a 300 lines mm$^{-1}$ grism blazed at 5000
\AA\ (2.64 \AA\, pix$^{-1}$ dispersion) in the blue channel.  To split the
red and blue channels, we used a dichroic with a cutoff at 6800 \AA.
Together, the two channels in this set--up afforded a spectral coverage of
roughly 3200 \AA\ to 1 micron.  Again, we observed the entire optical
window, but at almost twice the dispersion of our typical spectrograph
configuration.

We used the IRAF\footnote{IRAF is distributed by the National Optical
Astronomy Observatories, which are operated by the Association of
Universities for Research in Astronomy, Inc., under cooperative agreement
with the National Science Foundation.} package \citep{tody93} to process
both the longslit and the slitmask data, following standard slit
spectroscopy procedures.  Some aspects of treating the slitmask data were
facilitated by a home--grown software package, BOGUS\footnote{BOGUS is
available online at   
http://zwolfkinder.jpl.nasa.gov/$\sim$stern/homepage/bogus.html.}, created
by D. Stern, A.J.  Bunker, and S.A. Stanford.  Wavelength calibrations
were performed in the standard fashion using Hg, Ne, Ar, and Kr arc lamps;
we employed telluric sky lines to adjust the wavelength zero--point.  We
performed flux calibrations with longslit observations of standard stars
from \citet{massey90} taken with the instrument in the same configuration
as the relevant multislit observation.  However, it should be noted that
the absolute scale of the fluxed spectra must be regarded with caution.  
Not all of the nights were photometric; there may be substantial slit   
losses in the case of an extended source; small errors in slitmask
alignment cause additional light loss; and since the position angle of an
observation was set by the desire to maximize the number of targets on a
mask, the observations were in general not made at or near parallactic  
angle.  Moreover, in the case of serendipitous detections, it is unlikely
that the object is optimally aligned with the slit even when all other   
parameters are perfect.  Fortunately, it is merely the redshift of a given
object --- not the absolute flux or continuum shape --- which is of
interest at present.

\section{Visual Identifications and Redshift Determinations}
\label{ids}

\subsection{Visual Identifications}

A serendipitous detection in spectroscopic data presents two challenges to
the observer:  (1) to locate on an image of the field the progenitor of
the spectroscopic signature, and (2) to determine the nature of the object
and, where possible, its redshift.  We now address the former problem;  we
discuss the latter in the following section. 

In some respects, the process of cataloguing serendipitous detections
proceeds backwards from the usual steps involved in compiling redshifts. 
Generally, one begins with photometry for a galaxy whose location is known
and subsequently obtains a spectrum.  In our case, we began with a
spectrum and worked backward to the progenitor's location and photometry. 
To accomplish this task, we combined what was known about the observation
--- the location of the target, the dimensions and orientation of the
target slit, and the position of the target {\it within} the slit --- and
we reconstructed the position of the slit on the sky.  We then mapped the
reconstructed slit image to the target field and thereby identified {\it a
posteriori} the objects which we in fact observed. 

In the most favorable cases, the two--dimensional spectra contained
multiple serendipitous detections.  By comparing the relative spatial
separations between continuum detections in the two--dimensional spectra
to the separations between sources on the slit image, we could uniquely
identify each of the progenitors.  Even under unfavorable circumstance, in
which the target was too faint to appear in the spectrum or was
mis--aligned with the slit, the progenitor of a lone serendipitous
detection could be identified by comparing its position on the
two--dimensional spectrum to its position in the image relative to the
edges of the slit. 

To this end, the galaxies in the sample divide into two categories:  those
within the central HDF, and those without.  For galaxies inside of the
central HDF, we made the visual identification by mapping the slit image
to the remarkably deep, well--resolved central $I_{814}$ images presented
in W96.  We label the galaxies in Table \ref{hdfz} by their IDs, isophotal
magnitudes, and positions as given in that paper.  If, on the other hand,
the target slit extended outside of the central HDF, we relied on
supporting photometry provided by the single--orbit $I_{814}$ Flanking
Field images of W96, the deep Hawaii 2.2m $V$, $I$, $H+K$ images of
\citet[][hereafter B99]{barger99}, or the deep $U_n$, $G$, $\cal{R}$
images of \citet{steidel96b}.  Since there is no existing nomenclature for
sources beyond the central HDF, we adopted a labeling scheme in Table
\ref{ffz} based on the galaxy positions.

To facilitate this position--based nomenclature for serendipitous
observations of galaxies in the Flanking Fields and beyond, we computed an
astrometric solution to the Hawaii 2.2m $I$--band image of B99. From a fit
to 72 objects in a 10\arcmin\, $\times$ 10\arcmin\, portion of the
digitized POSS--II plates\footnote{ The Second Palomar Observatory Sky
Survey (POSS-II) was made by the California Institute of Technology with
funds from the National Science Foundation, the National Aeronautics and
Space Administration, the National Geographic Society, the Sloan
Foundation, the Samuel Oschin Foundation, and the Eastman Kodak
Corporation.} (obtained via the Digitized Sky Survey\footnote{ The
Digitized Sky Survey was produced at the Space Telescope Science Institute
under U.S. Government grant NAG W-2166. The images of these surveys are
based on photographic data obtained using the Oschin Schmidt Telescope on
Palomar Mountain and the UK Schmidt Telescope. The plates were processed
into the present compressed digital form with the permission of these
institutions.}), we found a platescale of 0\farcs189 pixel$^{-1}$ and an
orientation angle of 0.630\deg, both of which are consistent with the
values reported by B99.  The dispersion in the fit was 0\farcs22 in the
right ascension (RA) direction and 0\farcs26 in the declination (Dec) 
direction, giving a total error of 0\farcs37.  This error is comparable to
the error reported by B99.  As a check to the fit, we compared the RA and
Dec positions of 10 objects in the central HDF as given by W96 against our
newly computed Hawaii 2.2m $I$--band positions and found a mean absolute
offset of 0\farcs11 in RA and 0\farcs16 in Dec, for a total mean offset of
0\farcs20.  This error is smaller than the sum in quadrature of the total
dispersion in our fit and the accuracy of the W96 absolute astrometry
(reported as good to approximately 0\farcs4), suggesting that our reported
RA and Dec positions themselves are good to roughly 0\farcs4. 

In three cases, the serendipitous detection fell outside of the Hawaii
2.2m fields.  To visually identify these objects, we utilized our own 70
minute $R$--band image taken with the Echelle Spectrograph and Imager
\citep[ESI,][]{sheinis00} on UT 2000 May 05.  See \citet{stern00b} for a
detailed account of the observation and data reduction.  The astrometric
solution for the position--based nomenclature was determined exactly as
described for the Hawaii 2.2m image;  $I_{\mbox{\tiny AB}}$ magnitudes are
not available for these detections.  In five cases, the progenitor of a
serendipitous spectroscopic detection was too faint to be detected in any
of the supporting imaging.  Nonetheless, we were able to estimate a
position for the source by extrapolating from the known position of the
target and the dimensions and orientation of the target slit.  We have
indicated such cases on Table \ref{ffz}.

\subsection{Redshifts}
\label{redshifts}

For each member of the serendipitous catalog, we measured the redshift by
visually inspecting the spectrum and noting the wavelengths of spectral
features.  For objects with multiple strong emission lines, the proper
interpretation of the spectral features was unambiguous and the assignment
of their rest wavelengths was straightforward.  The situation was more
difficult for faint objects showing only absorption lines.  If such a
spectrum did not conform to the standard pattern of Balmer lines and the
H$+$K \ion{Ca}{2} doublet seen in the vicinity of the 4000 \AA--break (D4000),
then it was generally impossible to determine a redshift. 

The most common type of serendipitous detection involved the presence of a
single emission line, the interpretation of which can problematic.  In
general, a single, isolated line could be any one of Ly$\alpha$, \oii,
H$\beta$, \oiiib, or H$\alpha$, though given sufficient spectral coverage,
most erroneous interpretations can be ruled out.  For instance, the
absence of H$\beta$ or \oiiia\ serves to discount the interpretation of a
solo line as \oiiib.  Similarly, lines that are bluer than rest H$\alpha$
cannot be H$\alpha$ themselves, and the presence of H$\beta$ or \oiiib\
would be expected for a solo line redder than rest H$\alpha$ (but see
Stockton \& Ridgway 1998).  Hence, the primary threat to determining
one--line redshifts is the potential for mis--identifying Ly$\alpha$ as
\oii\ or vice versa.  Unfortunately, with low dispersion spectra it is
often impossible to distinguish between the high equivalent width forms of
these emission lines without a pronounced continuum depression or a line
asymmetry, both characteristic of Ly$\alpha$.  For two accounts of the
potential pitfalls associated with one--line spectroscopic redshift
identifications, see \citet{stern00a} and \citet{manning00}. 

\begin{figure}
\centering
\epsscale{0.7}
\plotone{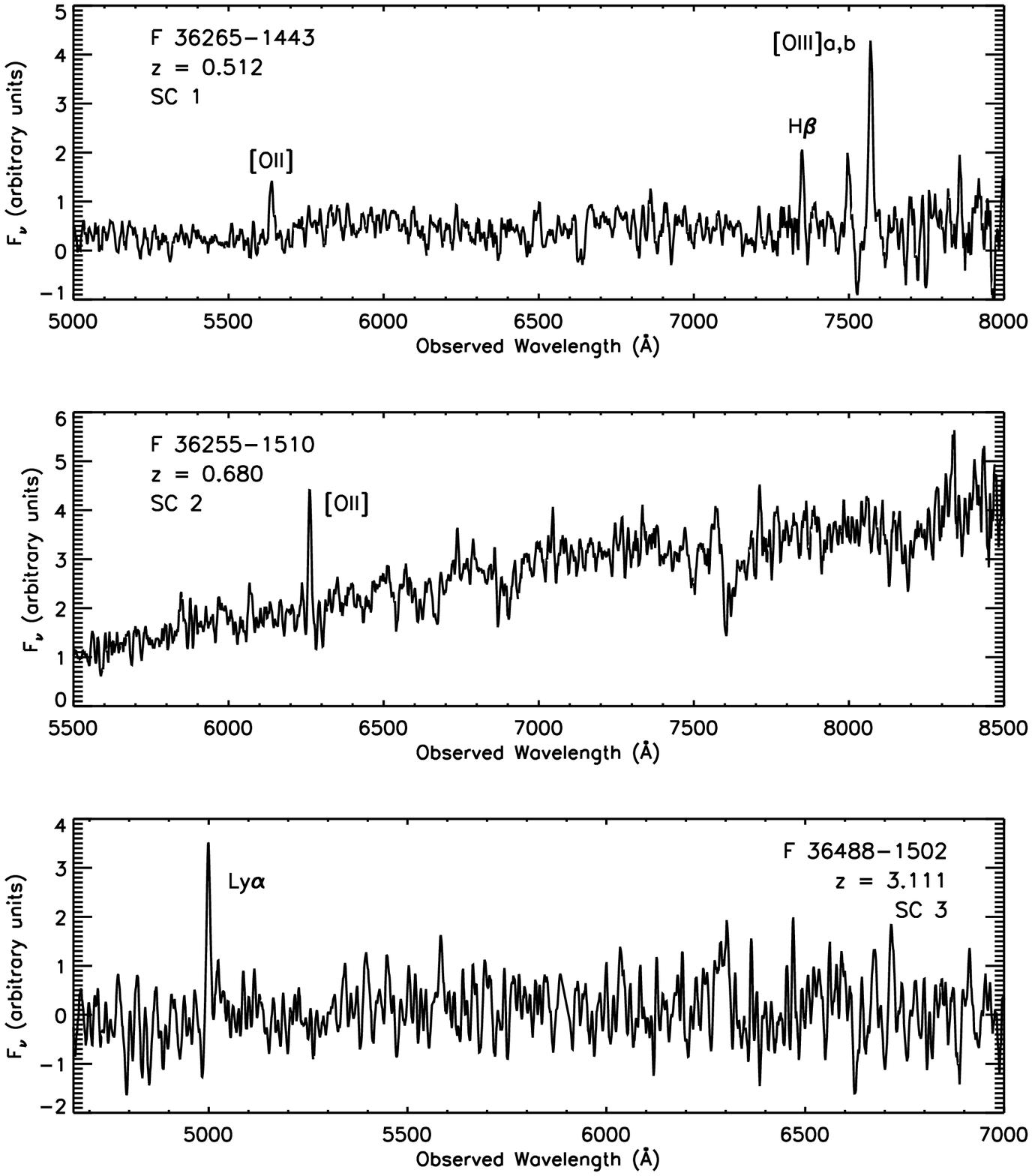}
\caption{Example spectra for spectral categories 1, 2, and 3.
See \S \ref{redshifts} and Table \ref{qc} for a detailed
account of the spectral categories.
The total exposure time for each is 5.4 ks.
The spectra have been smoothed using a 20 \AA\ boxcar filter.}
\label{QC1_3}
\end{figure}

In part to reflect the uncertainty in interpreting solo lines, we divide
the serendipitous detections into five spectral categories (SC) based on
their general morphology.  Table \ref{qc} lists the spectral categories
with a brief description of each.  The spectra of category 1 sources show
multiple features which can be uniquely identified, yielding secure
redshift determinations.  The spectra of category 2 sources show a solo
emission line in the presence of strong continuum both redward and
blueward of the line.  Such lines were identified as \oii, and the
redshift determination is considered secure.  The spectra of category 3
sources show a solo emission line redward of a strong continuum break. 
Such lines were identified as Ly$\alpha$ and the continuum breaks were
interpreted as the onset of absorption by the Ly$\alpha$--forest (which
causes significantly diminished flux shortward of 1216 \AA).  Of course,
especially in star--forming systems, the continuum in the vicinity \oii\
can also show a break --- the Balmer break at 4000 \AA\ --- and
in cases of low signal--to--noise, the morphology of the Balmer break
alone is not sufficient to distinguish it from the break at Ly$\alpha$. 
Fortunately, for galaxies at $z \simgtr 4$, the break at Ly$\alpha$ is
expected to be of greater amplitude than the largest observed D4000
amplitudes \citep[see][Fig.\ 12]{stern99}, so the two features can be
easily discerned.  At lower redshifts, however, the amplitude of the two
breaks may be comparable, and without corroborating spectral features the
redshift identification is largely subjective.  Of five category 3 sources
in this catalog, two are at $z \simgtr 4$, one has a redshift which is
confirmed by other authors, and one has supporting photometric redshifts
from two other authors; their redshift determinations are considered
secure.  The redshift of the remaining category 3 source should be
considered tentative, as indicated on Table \ref{ffz}.  Example spectra
for categories 1, 2, and 3 are shown in Figure \ref{QC1_3}. 

\begin{figure}
\centering
\plotone{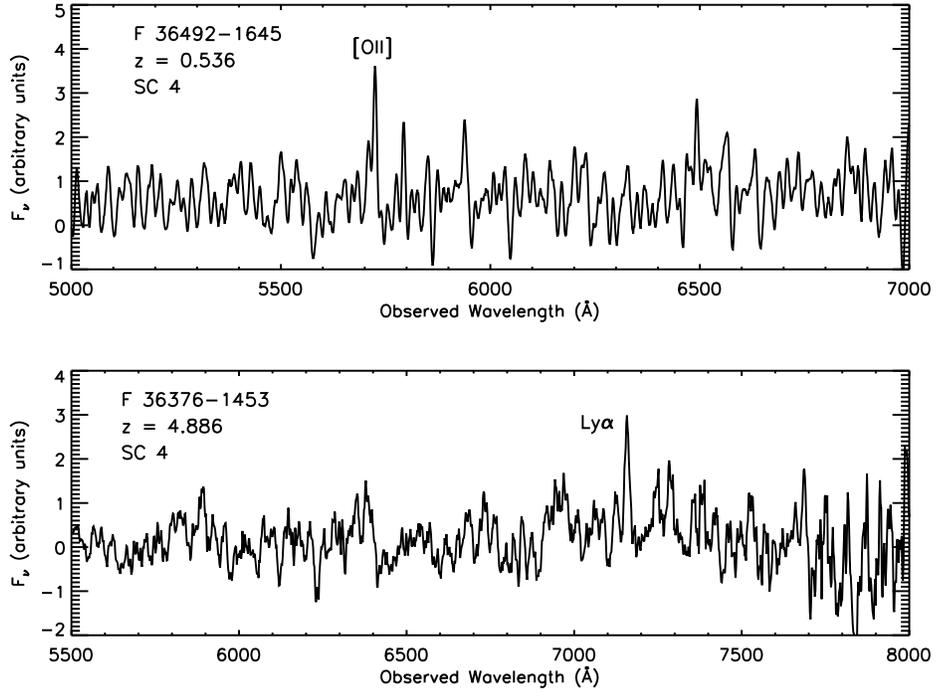}
\caption{Example spectra for spectral category 4, in which
a solo emission line in the absence of continuum
is identified as either \oii\
(top panel) or as Ly$\alpha$ (bottom panel), based in part on the line profile,
the line observed--frame equivalent width, and/or the supporting imaging.
See \S \ref{redshifts} and Table \ref{qc} for a detailed account
of the spectral categories.
The total exposure time for each is 5.4 ks.
The spurious features observed in the continuum are
due to residuals from the subtraction of strong telluric
OH emission lines.  The blueward ``shoulder" on the
\oii\ emission line is an imperfectly removed
cosmic ray.
The spectra have been smoothed using a 20 \AA\ boxcar filter.}
\label{QC4}
\end{figure}

\begin{figure}
\centering
\plotone{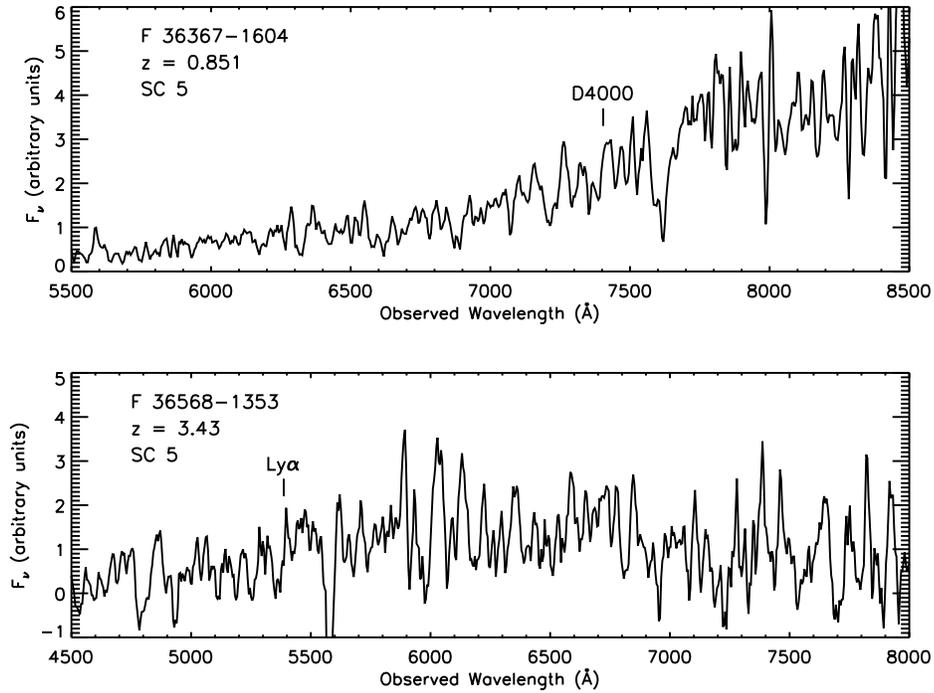}
\caption{Example spectra for spectral category 5, in which
a continuum break is interpreted as the 4000 \AA--break (top panel) or
as the onset of Ly$\alpha$--forest absorption (bottom panel)
according the strength of the
continuum blueward of the break.
See \S \ref{redshifts} and Table \ref{qc} for a detailed account
of the spectral categories.
The total exposure time for each is 5.4 ks.
The spurious features observed in the continuum are
due to residuals from the subtraction of strong telluric
OH emission lines.
The spectra have been smoothed using a 20 \AA\ boxcar filter.}
\label{QC5}
\end{figure}

The spectra of category 4 objects show an isolated emission in the absence
of any continuum, which generally suggests a weak detection of either
\oii\ or Ly$\alpha$.  Clearly, the confidence one can exercise in
discriminating between these two cases is a strong function of the
robustness of the detection, the resolution of the spectrum, and the
availability of supporting imaging.  See \S \ref{highz} for a detailed
discussion of the redshift determination of a typical category 4 source. 
Example spectra for both interpretations of category 4 sources are shown
in Figure \ref{QC4}.  The spectra of category 5 sources show a continuum
break.  Such breaks were classified as either the Balmer break or as
Ly$\alpha$--forest absorption according to the strength of the continuum
blueward of the break.  Example spectra for both interpretations of
category 5 are shown in Figure \ref{QC5}.  The redshift determinations of
both category 4 and category 5 sources are considered secure unless
otherwise indicated.  Serendipitous detections about which we were unable
to attain a reasonable degree of confidence were omitted from the catalog;
nearly half of the initial sample of 121 serendipitous detections were
rejected for this reason. 

To minimize the possibility that we mis--classified the solo emission line
of a low--redshift category 4 source as high--redshift Ly$\alpha$, we
checked that the source as visually identified in the Hawaii 2.2m
$I$--band image of B99 did not also appear in the $\cal{R}$--band image of
\citet{steidel96b}.  In this fashion, we ensured that the $\cal{R}$--band
flux of the source in question was severely attenuated by the hydrogen
forest, consistent with $z \simgtr 4$.  We discovered one erroneous
redshift determination with this technique:  F 36265--1443 was marginally
detected in 1999 June such that \oiiib\ appeared in the two--dimensional
spectrum as a solo emission line at $\lambda = 8136$ \AA, and the line was
initially mis--classified as high--redshift Ly$\alpha$ at $z = 5.691$. 
However, the presence of the progenitor in the $R$--band image of
\citet{steidel96b} ruled out the tantalizing high--redshift
interpretation, and subsequent targeted spectroscopy revealed a spectrum
with \oii, \oiiia, \oiiib, H$\beta$, and H$\gamma$ in emission at
$z=0.625$. 

In the event that a redshift for a serendipitous detection remained
undetermined, one possibility is that the object lies in the so--called
``redshift desert,"  the interval spanning roughly $1.7 < z < 2.3$.  The
limits of this interval are set by the fact that at higher redshifts
Ly$\alpha$ would fall on the detector, and at lower redshifts the oxygen
lines and/or the Balmer lines would fall on on the detector.  At
intermediate redshifts, however, there is a dearth of prominent spectral
features, rendering redshift determination difficult.  A second
possibility is that the object does have spectral features which are in
principle observable, except that the features fall in a region heavily
contaminated by night sky emission.  As sky subtraction is particularly
problematic at $\lambda > 7200$ \AA\ for low signal--to--noise, low
dispersion spectra, it is reasonable to conclude that at least a few
redshifts were lost to this effect. 

It should be noted that for the $\sim 5$ cases in which a single galaxy
was multiply observed, the agreement in the individual redshifts was
excellent.  Discrepancies never exceeded $\Delta z = 0.004$. 

\section{The Catalogs}
\label{catalogues}

We present the catalog of serendipitously detected galaxies in Table
\ref{hdfz} and Table \ref{ffz}.  Table \ref{hdfz} contains 25 galaxies
located in the HDF proper, identified by their W96 number as described in
the preceding section.  The $I_{814}$ magnitude is the isophotal magnitude
given by W96, and the RA and Dec are J2000 coordinates, also given
therein.  The spectral category was assigned as described in \S 3.1; also
see Table \ref{qc}.  Table \ref{ffz} contains 49 galaxies located outside
the central HDF, identified by their positions as described in the
preceding section.  The 30 galaxies located in the HDF Flanking Fields are
indicated.  The isophotal $I_{\mbox{\tiny AB}}$ magnitudes were determined
by running the source extraction algorithm SExtractor \citep{bertin96} on
the Hawaii 2.2m $I$--band image of B99.  We estimated the $I_{\mbox{\tiny
AB}}$ zero--point by using stars in the central HDF; as such, the
uncertainty in the $I_{\mbox{\tiny AB}}$ is $\sim 0.3$ mag.  All spectral
lines in both tables are emission lines unless otherwise noted. 
 
\section{Discussion}
\label{discuss}
 
The 74 galaxies in the serendipitous catalog span the redshift range $0.10
< z < 5.77$, with a median redshift of $z = 0.85$.  The vast majority of
the galaxies are emission-line systems; 5\% of the sample show only
absorption lines.  This bias stems from the diminished likelihood of
serendipitously detecting an absorption--line system with sufficient
signal--to--noise to allow the redshift to be determined.

We estimate that the uncertainty in the most secure redshifts (SC 1) is $|
\Delta z | \approx 0.003$.  The uncertainty in redshifts based on solo
emission lines or continuum breaks (SC 2 to 5) --- assuming the
identification of the spectral feature is sound --- is $| \Delta z |
\approx 0.004$.  For the 12 galaxies in the central HDF also observed
spectroscopically by \citet{cohen96}, \citet{cohen00}, \citet{phillips97},
or \citet{steidel96a}, we compared our spectroscopic redshift to the
published value and found that the agreement was excellent, with a mean
deviation of $| \Delta z | = 0.001$ and a dispersion of $\sigma_{\Delta z}
= 0.001$.  In all cases, the discrepancy is comparable to our estimated
measurement error. 

\subsection{The Redshift Distribution}

\begin{figure}
\centering
\epsscale{0.7}
\plotone{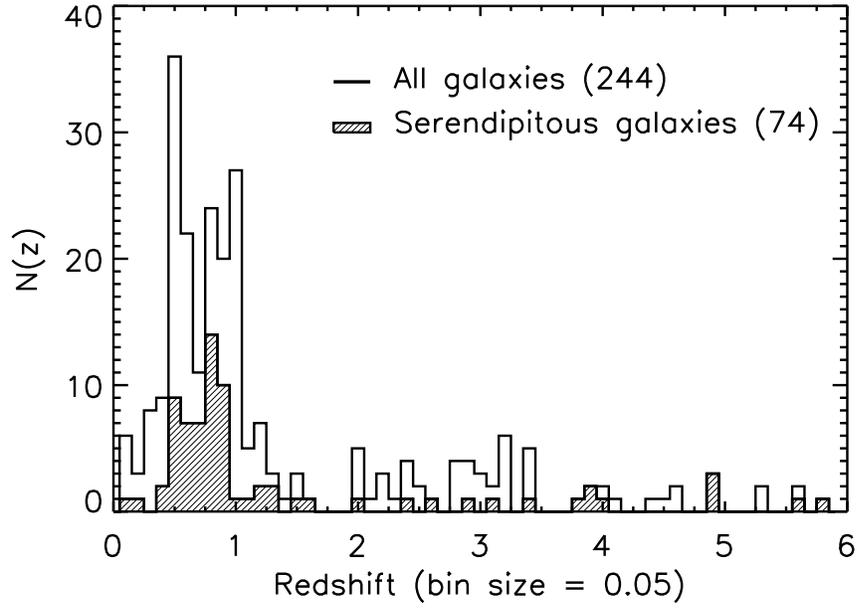}
\caption{
Distribution of redshifts of the serendipitous sample compared to a total
sample consisting of all published redshifts for galaxies in the central
HDF plus 26 galaxies which flank the central HDF.
\label{histo_full}}
\end{figure}

\begin{figure}
\centering
\epsscale{0.7}
\plotone{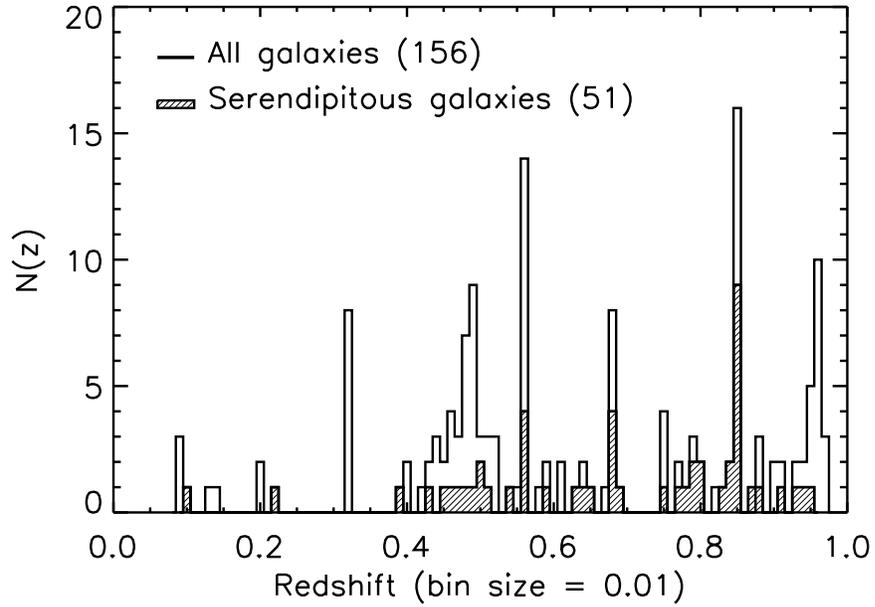}
\caption{
Distribution of redshifts of the serendipitous sample compared to the
total sample for the range $0 < z < 1$.
\label{histo_short}}
\end{figure}

The redshift distribution of the serendipitous catalog, compared with a
``total sample" consisting of this sample, all published redshifts for
galaxies in the central HDF, and 26 published redshifts for galaxies
flanking the central HDF, is shown in Figures \ref{histo_full} and
\ref{histo_short}.  Sources for the total sample are \citet{bunker98}; 
\citet{cohen96};  \citet{cohen00};  \citet{lowenthal97}; 
\citet{phillips97};  \citet{spinrad98};  \citet{stern99}; 
\citet{waddington99}; and \citet{weymann98}.  The histogram displayed in
Figure \ref{histo_full} displays the total range of redshifts of the
combined catalogs, $0.089 < z < 5.77$, with a comparatively coarse
resolution of $\Delta z = 0.1$.  Given the caveat that we are insensitive
to galaxies in the redshift range $1.7 < z < 2.3$ (cf.\ \S3.1), we find
that the redshift distribution of the serendipitous sample closely follows
that of the total sample.

To investigate the redshift clustering properties of the serendipitous
sample, we display the redshift distribution for the galaxies in the range
$0 < z < 1$ with a resolution of $\Delta z = 0.01$ in Figure
\ref{histo_short}.  The figure shows clear evidence of clustering in both
the serendipitous sample and the total sample.  Moreover, the clustering
present in the total sample is mirrored almost perfectly by that present
in the serendipitous sample.  Assuming a fixed number of galaxies per
redshift bin (i.e.\ no evolution in bin membership with redshift), we find
a $2.3 \sigma$ peak in the serendipitous sample at $z = 0.79$, a $3.2
\sigma$ peak at $z = 0.56$ and $z = 0.68$, and a $6.9 \sigma$ peak at $z =
0.85$. In total, we find that 17 out of the 51 serendipitous galaxies
(33\%) fall into peaks significant at greater than 97.5\% confidence. 
This figure compares favorably with that of \citet[][hereafter
C96]{cohen96}, who find that 57 out of 140 (41\%) of their sample of
spectroscopically observed HDF galaxies fall into redshift peaks.  That
the locations of our peaks vary somewhat from those in C96 is not
surprising.  Whereas C96 chose redshift bins of variable centers and
widths so as to maximize their significance relative to occurring by
chance in a smoothed velocity distribution, we chose fixed bin centers and
widths, cf.\ \citet{phillips97}.  Even so, our peaks centered on $z =
0.56$ and $z = 0.68$ no doubt reflect the same structures revealed by the
peaks in C96 at $z_p = 0.559$ and $z_p = 0.680$, respectively.  We find no
evidence of periodicity in the peak redshifts, as described by
\citet{broadhurst90}.

Beyond the strong evidence of redshift clustering, there are two
outstanding features of the redshift distribution of the serendipitous
sample.  First, there is a relative deficiency of serendipitous detections
at $z < 0.4$.  Second, the redshift peak centered on $z = 0.32$ evident in
the total sample is not represented in the serendipitous sample.  Taken
together, these features appear to suggest a selection effect which
excludes galaxies at $z < 0.4$ from serendipitous detection.  However,
since this redshift range is perfectly accessible to LRIS via the Balmer
lines and by [\ion{O}{2}] and [\ion{O}{3}] emission, it is likely that the
scarcity of low--redshift galaxies in the serendipitous catalog is merely
the combined effect of: (1) the increasingly small cosmological volume
surveyed at low--redshift, (2) the comparatively small size of the
serendipitous catalog, and (3) the fact that the HDF was selected to be
devoid of bright galaxies in the first place.  At a minimum, these facts
make it impossible to comment on the significance of the apparent $z <
0.4$ deficiency.

\subsection{A Check of Photometric Redshifts}

Photometric redshift techniques have become an essential tool of
observational cosmology, with applications ranging from determining
luminosity functions to selecting high--redshift candidates for
spectroscopy.  We have utilized our set of spectroscopic redshifts for 23
of the 25 serendipitously detected galaxies in the central HDF to carry
out a test of the photometric redshifts presented by \citet{fernandez99},
who employ a maximum--likelihood analysis applied to spectral energy
distribution--fitting of precise $U_{300}$, $B_{450}$, $V_{606}$,
$I_{814}$, $J$ (1.2 $\mu$m), $H$ (1.65 $\mu$m), and $K$ (2.2 $\mu$m)
photometry.  For two galaxies, HDF 4--402.1 and HDF 4--236.0, no
photometric redshift was available, no doubt owing to their faintness:
$I_{\mbox{\tiny AB}} = 24.96$ and 28.26, respectively.  The sample of
predicted redshifts was taken from the group's world wide web site --- the
University of New South Wales/State University of New York at Stony Brook
HDF Clickable
Map\footnote{http://bat.phys.unsw.edu.au/$\sim$fsoto/hdf/hdf\_fs.html} ---
which is an interactive version of the catalog presented in the associated
paper.

\begin{figure}
\centering
\epsscale{0.5}
\plotone{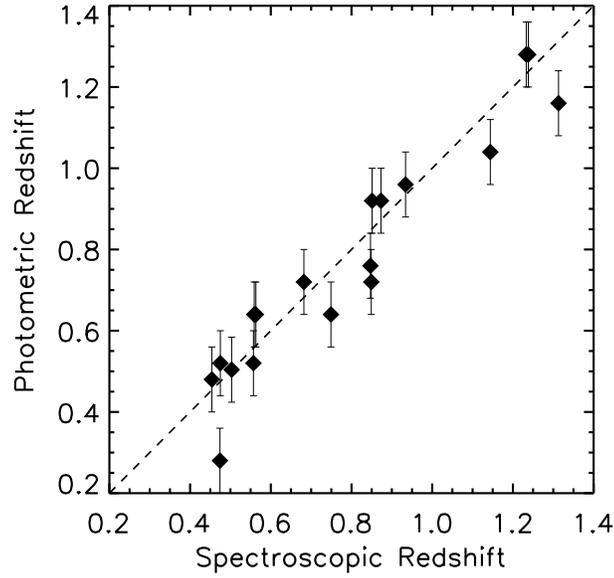}
\caption{Comparison of spectroscopic and photometric redshifts
for 17 serendipitously detected galaxies in the central HDF.
The error bars are attributable to cosmic variance,
the fact that the model spectra used in determining
photometric redshifts represent a finite sample of all possible
galaxy spectra; photometric errors are negligible in
this redshift range.  Three obviously
erroneous photometric redshifts are off the scale:
HDF 4--639.1, listed with
$\zs = 2.592$ and $\zp = 0.000$;
HDF 2--600.0, listed with $\zs = 0.425$ and $\zp = 1.800$; and
HDF 4--658.0,
listed with $\zs = 0.558$ and $\zp = 4.320$.
\label{spec_vs_phot}}
\end{figure}

We compare the spectroscopic redshift ($\zs$) and the photometric redshift
($\zp$) in a scatter plot of $\zs$ versus $\zp$ for redshifts less than   
1.5 in Figure \ref{spec_vs_phot}.  There are three obvious errors in the
photometric redshifts: (1) HDF 4--639.1, listed with $\zs = 2.592$ and
$\zp = 0.000$, whose spectrum shows Ly$\alpha$ in emission with a strong
continuum break (SC 3), and whose $\zs$ is confirmed by both
\citet{steidel96a} and \cite{cohen00};  (2) HDF 2--600.0, listed with $\zs
= 0.425$ and $\zp = 1.800$, whose spectrum shows a strong solo emission
line interpreted as \oii\ (SC 4); and (3) HDF 4--658.0, listed with $\zs =
0.558$ and $\zp = 4.320$, whose spectrum shows both [\ion{O}{2}] and   
[\ion{O}{3}] emission (SC 1).  These outliers comprise 13\% of the sample,
roughly consistent with the finding of \citet{cohen00} that outliers at 
more than 4$\sigma$ in the $\zs$--$\zp$ plane comprised $\sim 10$\% of the
subset of galaxies at $z < 1.5$. The outliers are not shown in Figure   
\ref{spec_vs_phot}, as they are off the scale.

The mean and the dispersion of the difference between the predicted
photometric redshifts and the measured spectroscopic redshifts are $|
\Delta z | = 0.380$ and $\sigma_{\Delta z} = 0.907$, respectively. 
However, these values are dominated by the three discrepant points
described above.  When the discrepant points are omitted, we find a mean
deviation of $| \Delta z | = 0.098$ and a dispersion of $\sigma_{\Delta z}
= 0.010$.  These errors are consistent with the assessment that cosmic
variance (the fact that the model spectra used in determining photometric
redshifts represent a finite sample of all possible galaxy spectra) rather
than photometric errors is the dominant source of error at small redshift
\citep{fernandez99}.  Moreover, these results confirm that --- barring
catastrophic errors --- photometric redshifts are capable of producing
reasonable predictions of galaxy redshifts where suitably precise
multicolor photometry is available.

\subsection{A Galaxy Cluster at $z = 0.85$}

We report the serendipitous discovery of ClG 1236+6215, a galaxy cluster
with redshift $z = 0.85$ nominally centered at
$\alpha=$12\hr36\min39\secper6,
$\delta=+$62\deg15\arcmin54$^{\prime\prime}$ (J2000).  The cluster was
initially identified as an over--density of centrally concentrated red
objects in a small region to the northwest of the HDF in the deep Hawaii
2.2m $V$ and $I$ images of \citet{barger99}.  In a circle of radius 45
arcsec centered on the cluster position, the density of objects with
$(V-I)_{\mbox{\tiny AB}} > 1.5$ is 18 arcmin$^{-2}$, versus a density of
only 6.5 arcmin$^{-2}$ over the rest of the 90 arcmin$^2$ Hawaii 2.2m
field.  We interpreted the $(V-I)_{\mbox{\tiny AB}}$ color of the
concentration to be the result of the 4000 \AA\ break redshifted into the
$I$--band, and we targeted five of the reddest members for spectroscopy. 
All five of the targets proved to have redshifts very near to $z = 0.85$. 
We added three more redshifts by selecting objects from the redshift
catalog of \citet{cohen00} which had $(V-I)_{\mbox{\tiny AB}} > 1.5$ and
$0.84 < z < 0.86$, and which were located within 45 arcsec ($0.17 \,
h_{100}^{-1}$ Mpc) of the cluster center.  Together, the eight
spectroscopic members of ClG 1236+6215 yield a mean redshift for the
cluster of $z = 0.849 \pm 0.004$.  The properties of the spectroscopic
members of ClG 1236+6215 are summarized in Table \ref{clg}. 

Following the prescription of \citet{harrison74} for properly considering
the contributions to measured redshifts due to the radial component of the
motion of our Galaxy with respect to the Local Group, to the cosmological
expansion between comoving observers at our Galaxy and at the galaxy
cluster, and to the radial component of the peculiar velocity of the
galaxy within the cluster, we calculated an estimate of the corrected
line--of--sight velocity dispersion \sig\ in ClG 1236+6215.  We followed
the treatment of \citet{danese80} to account for the spurious systematic
contribution to \sig\ from measurement errors in the member redshifts. 
Assuming an underlying Gaussian distribution for the galaxy velocities, we
found $\sig = 610 \pm 190$ km s$^{-1}$ (68\% confidence);  this value
should be treated with caution due to the small number of spectroscopic
members.  \citet{beers90} point out that the classical standard deviation
estimator for cluster velocity dispersions is neither resistant to the
presence of outliers nor robust for non--Gaussian underlying populations. 
However, employing the ``gapper" method as implemented in their ROSTAT
package yields a correction which is less than our estimated uncertainty.

In the limiting isothermal model, the calculated velocity dispersion
translates to a mean cluster mass within a 45 arcsec ($0.17 \,
h_{100}^{-1}$ Mpc)  radius of the cluster center of $M(r<0.17 \,
h_{100}^{-1} \mbox{ Mpc}) = 2.6 \times 10^{13} \, h_{100}^{-1} \,
M_{\odot}$.  For comparison with other authors, the mean mass within the
Abell radius is $M(r<1.5\, h_{100}^{-1} \mbox{ Mpc}) = 2.3 \times 10^{14}
\, h_{100}^{-1} \, M_{\odot}$.  Of perhaps more immediate observational
consequence is the X--ray luminosity expected for the given velocity
dispersion.  Drawing on a sample of 197 galaxy clusters --- which
constitutes the largest cluster data set used to date for such a study ---
\citet{xue00} find $L_X / (10^{43} \mbox{ erg s}^{-1}) = 10^{-13.68 \pm
0.61} \sig^{5.30 \pm 0.21}$ for the X--ray bolometric luminosity--velocity
dispersion relation.  This result yields an expected X--ray bolometric
luminosity for ClG 1236+6215 of $L_X = 1.2 \times 10^{44}$ erg s$^{-1}$, a
value which exceeds the expected detection threshold of the upcoming
$\approx 1$ Ms {\em Chandra X--ray Observatory} (CXO) exposure of the HDF and
its environs \citep{brandt01}. 

\subsection{Optical Spectroscopy of the X--ray Source CXOHDFN J123635.6$+$621424}
\label{chandra}

Optical spectroscopy of faint X--ray sources is the key to determining the
poorly understood physical properties of the population responsible for
producing the X--ray background.  We present the first published optical
spectrum and redshift for CXOHDFN J123635.6$+$621424, a well--observed
X--ray source identified with a face--on spiral galaxy at $z = 2.011$,
fortuitously located in the Inner West HDF Flanking Field.

CXOHDFN J123635.6$+$621424 was first detected as a weak radio source (8.15
$\mu$Jy at 8.5 GHz; 87.8 $\mu$Jy at 1.4 GHz) in the sensitive HDF radio
surveys of \citet{richards98, richards00}.  The source has a comparatively
steep radio spectral index ($S_{\nu} \propto \nu^{-\alpha}$;
$\alpha^{\mbox{\tiny 8.4 GHz}}_{\mbox{\tiny 1.4 GHz}} > 0.87$), and the
radio emission extends across 2\farcs8.  In general, microjansky radio
emission from disk galaxies can result from either star formation (e.g.\
from free--free emission originating in \ion{H}{2} regions)  or from AGN
activity connected with a central engine.  \citet{richards98, richards00}
argued that (1) in the case of a central AGN powering a weak ($P <
10^{25}$ W Hz$^{-1}$)  radio source, the bulk of the radio emission is
confined to the nuclear region and is therefore characterized by
sub--arcsecond angular scales, and (2) such small scales result in a high
opacity to synchrotron self--absorption, yielding flat or inverted
spectral indices typically in the range $-0.5 < \alpha < 0.5$.  Hence, the
origin of the radio emission in CXOHDFN J123635.6$+$621424 was taken to be
extended star--forming regions.  This conclusion was ostensibly borne out
by an {\em Infrared Space Observatory} Camera (ISOCAM) detection of the
source \citep{aussel99}.  If the source were a moderate--to--low redshift
starburst galaxy (as suggested by Hornschemeier 2001, owing to the
object's spatial extent), the ISOCAM 15 $\mu$m filter (LW3) would sample
rest wavelengths from roughly 6 $\mu$m to 12 $\mu$m; the mid--infrared
emission could therefore be plausibly attributed to the unidentified
infrared bands (UIB) and to the hot, 200 K dust which typically dominates
the spectral energy distribution of starbursts over those wavelengths
\citep{aussel99}.

\begin{figure}
\centering
\epsscale{0.8}
\plotone{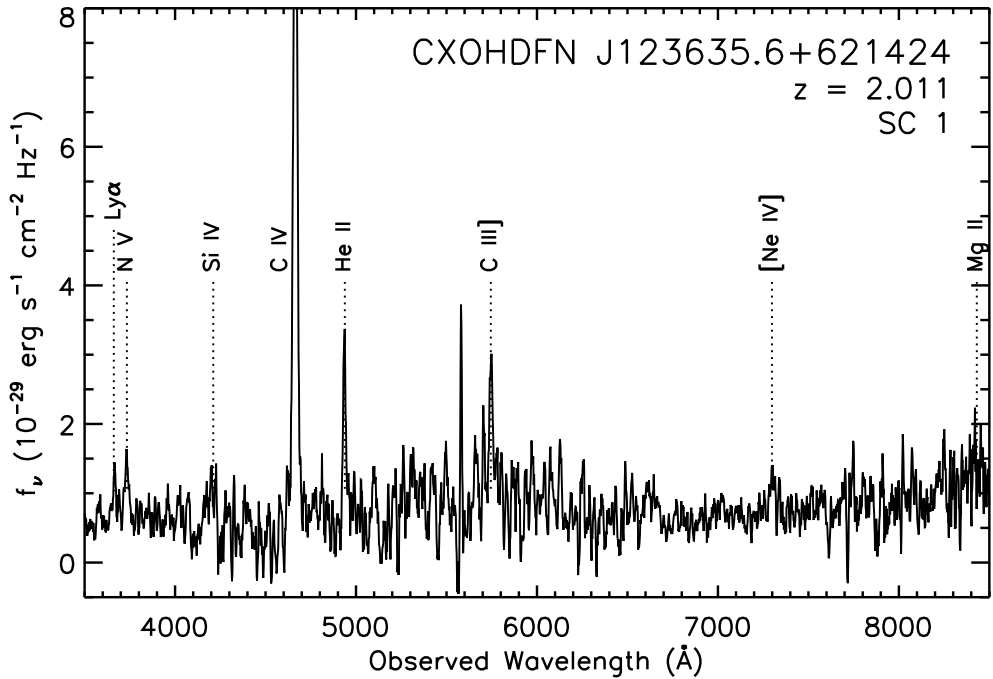}
\caption{Optical spectrum of the X--ray source CXOHDFN J123635.6$+$621424.
The spectrum was obtained on UT 24 February 2001, after the advent of the
LRIS--B spectrograph channel.  Flatfield and flux--calibration
difficulties associated with the blue channel
prevented us from calibrating the blue side ($\lambda < 6800$ \AA)
in the standard fashion.  To create the spectrum shown, 
we assumed a flat rest UV spectrum ($f_{\nu} \propto \nu^0$)
and then forced the blue channel and red channel fluxes to agree at 6800 \AA.
Though line ratios determined within either spectrograph channel (e.g.\
the $\lambda$1549/\ion{He}{2} $\lambda$1640 ratio of $\sim 8$ cited
in \S \ref{chandra}) are reliable, line ratios made {\it across} spectrograph
channels should be considered suspect.
The total exposure time is 8.4 ks.  The spectrum was smoothed
using a $\sim 10$ \AA\ boxcar filter.}
\label{chandra_spec}
\end{figure}

In contradistinction to the foregoing conclusions, both the optical and
X--ray properties of CXOHDFN J123635.6$+$621424 indicate the presence of
AGN activity.  The optical spectrum shows moderate--width ($\sim 1000$   
km/s), high--ionization emission lines, similar to those of the recently
reported quasar II in the Chandra Deep Field South \citep{norman01} and 
typical of high--redshift radio galaxies \citep[cf.][]{mccarthy93,
stern99}.  We detect Ly$\alpha$, \ion{N}{5} $\lambda$1240, \ion{Si}{4}
$\lambda$1397, \ion{C}{4} $\lambda$1549, \ion{He}{2} $\lambda$1640,
\ion{C}{3}] $\lambda$1909, [\ion{Ne}{4}] $\lambda$2424, and \ion{Mg}{2} 
$\lambda$2800 (Figure \ref{chandra_spec}).  Moreover, the rest frame  
equivalent widths of the \ion{C}{3}] $\lambda$1908 and \ion{C}{4}
$\lambda$1548 emission lines ($\sim 13$ \AA\ and $\sim 100$ \AA,
respectively)  are within the ranges found in multiple AGN emission line
surveys and optical/radio quasar surveys \citep[see][and references
therein]{lehmann00}.  We also note that the \ion{C}{4}
$\lambda$1549/\ion{He}{2} $\lambda$1640 ratio of $\sim 8$ is more typical
of quasars than of radio galaxies.  Optical and near--IR photometry of   
CXOHDFN J123635.6$+$621424 corroborates these findings.  \citet{hogg00}
give $({\cal R} - K_s) = 4.74$ for the source, and \citet{hasinger99}
report that all X--ray counterparts with $(R - K^{\prime}) > 4.5$ in their
ROSAT Ultra Deep HRI Survey are either members of high redshift clusters  
or are obscured AGN.  Finally, CXO observations of the source indicate a
comparatively hard X--ray spectrum --- the definitive signature of an AGN.
The X--ray band ratio, defined as the ratio of hard--band (2 keV to 8 keV)
to soft--band (0.5 keV to 2 keV) number counts, is $0.75^{+0.71}_{-0.43}$,
corresponding to a photon index\footnote{The photon index $\Gamma$ is
derived from a power law model for the X--ray spectrum:  $N =
AE^{-\Gamma}$, where $N$ is the number of photons s$^{-1}$ cm$^{-2}$
keV$^{-1}$ and $A$ is a normalization constant.} of $\Gamma = 0.75$
\citep{hornschemeier01}.

When re--interpreted in the light of the spectroscopic redshift, even the
mid--IR data for CXOHDFN J123635.6$+$621424 actually indicate the
presence of an AGN.  For the derived redshift of $z = 2.011$, the ISOCAM
LW3 filter samples rest wavelengths spanning only 4 $\mu$m to 5 $\mu$m.
Here, the contribution to the mid--IR spectral energy distribution made by
UIB emission and by dust at 200 K is severely attenuated
\citep[see][Figure 1]{aussel99}.  Hence, the ISOCAM detection of this
source is far more plausibly explained by the hot, $\sim 10^3$ K dust
found in the central region of an AGN \citep[e.g.\ see][]{aussel98} rather
than by star formation alone.  The weakness of Ly$\alpha$ in this galaxy
substantiates the presence of dust in this system.

Though the canonical wisdom regarding extended radio sources with spectral
indices steeper than $\alpha > 0.5$ dictates that such sources are driven
by starbursts \citep{richards98,richards00,hornschemeier01}, the combined
weight of evidence from X--ray, optical, and near-- and mid--IR
observations of CXOHDFN J123635.6$+$621424 is definitively in favor of an
obscured AGN.  This conclusion is consistent with the trend reported by
\citet{hornschemeier01}:  that the high X--ray luminosities and large band
ratios of several CXO--detected radio sources previously reported as
starburst--type objects strongly suggests the presence of heretofore
unidentified AGNs.  We are currently pursuing Keck/NIRSPEC spectroscopy of
this interesting source in order to further detail its physical
properties.

\subsection{Galaxies at $z \stackrel{>}{\sim} 5$}
\label{highz}

In the course of deep, targeted spectroscopy of photometric high--redshift
galaxy candidates, we have identified several serendipitous high--redshift
Ly$\alpha$--emitting candidates, including five sources at $z \simgtr 5$. 
These high--redshift sources are evident in Figure \ref{histo_full}, and
they are listed in Table \ref{ffz}.  The surface density of such sources
is sufficiently high that these discoveries are not unexpected
\citep[\eg][]{dey98, manning00}.  Indeed, slit spectroscopy surveys for
high--redshift Ly$\alpha$ emission are fully complimentary to narrow--band
searches \citep[\eg][]{hu98, steidel99, rhoads99}: rather than probing a
large area of sky for objects over a limited range of redshift, deep slit
spectroscopy surveys a small area of sky for objects over a large range in
redshift \citep{pritchet94, thompson95}.  The total area covered by the
spectroscopic slits during the course of our study was $\approx 2.2$
arcmin$^{2}$, implying a surface density of $\approx 2.3$ arcmin$^{-2}$
Ly$\alpha$--emitters at redshift $z \sim 5$.  This value is roughly
consistent with the surface density of high--redshift Ly$\alpha$--emitters
reported by \citet{cowie98}: $\approx 3$ arcmin$^{-2}$ (unit--$z$)$^{-1}$
at redshift $z \sim 3.4$, for comparable sensitivity to line flux.  Of
course, one should exercise caution regarding these values, owing to the
small number of detections involved. 

Each of the high--redshift sources in this catalog are solo emission line
sources (SC 4), and as indicated by a handful of cautionary tales
\citep[\S 3.2 herein; also see ][]{stockton98,stern00a}, such redshift
identifications should be greeted with a degree of circumspection.  A
detailed discussion of each individual source is beyond the scope of this
paper, and a separate manuscript is planned.  For now, we restrict the
discussion to one likely high--redshift source, F 36246--1511 at $z =
5.631$, as illustrative of the situation. 

\begin{figure}
\centering
\includegraphics[angle=90,width=6in]{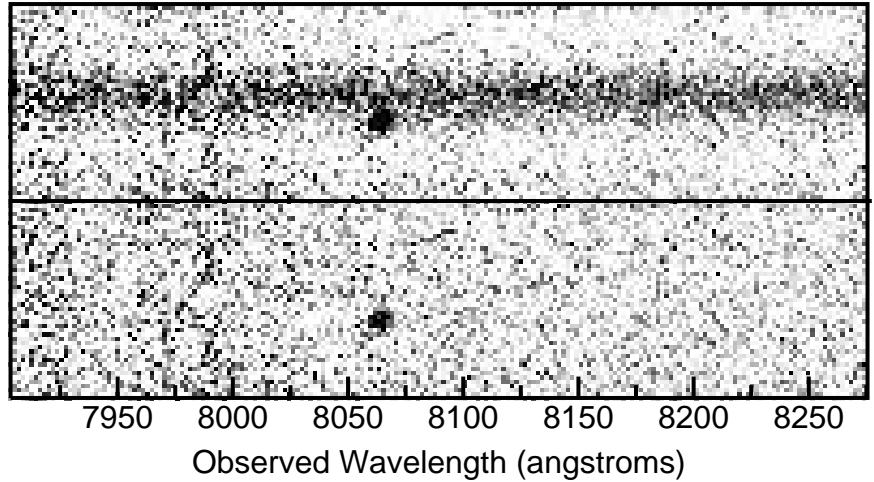}
\caption{Discovery spectrogram for F 36246--1511, a solo
emission line source interpreted as a Ly$\alpha$--emitter
at $z = 5.631$, lensed by an absorption--line
galaxy (F 36247--1510) at $z = 0.641$.
The top panel shows the raw spectrogram.
The bottom panel shows the solo emission line
after subtracting a Gaussian fit to the foreground continuum source.
The fit was made to the continuum source blueward
of the emission line so that after subtraction --- assuming a locally flat
spectrum for both sources --- any remaining flux could be attributed
to the solo line--emitter.
The total exposure time is 5.4 ks.
Each panel is 9\farcs5 in height.
Since the discovery spectrum was obtained, we have targeted F 36246--1511
for an additional $\sim$ 25 ks of spectroscopy.  The resulting
composite spectrum, which confirms the high--redshift interpretation,
will appear in a future work.
}
\label{serx2}
\end{figure}

\begin{figure}
\centering
\epsscale{0.6}
\plotone{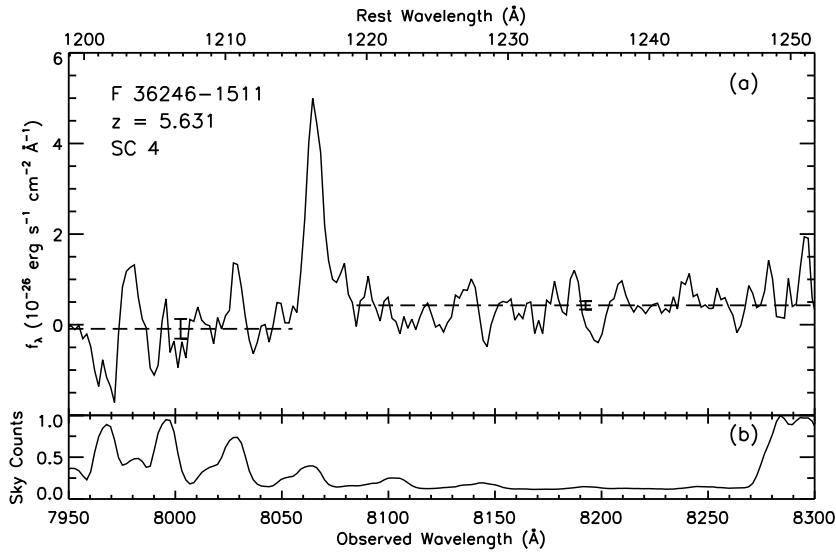}
\caption{(a) The one--dimensional extracted spectrum of F 36246--1511.
The dashed lines indicate the mean value of the spectrum over
wavelengths lower than and higher than the emission line.  The 
1--sigma scatter in the two regions is $0.7 \times 10^{-26}$
erg s$^{-1}$ cm$^{-2}$ \AA$^{-1}$ and $0.4 \times 10^{-26}$
erg s$^{-1}$ cm$^{-2}$ \AA$^{-1}$, respectively.  The error
bars indicate the 1--sigma scatter divided by the square root
of the number of resolution elements in each region.  Of course,
the meaningfulness of these statistics is contingent on the source
having a flat (or no) continuum on either side of the emission line.
The total exposure time is 5.4 ks.
The spectrum was smoothed using $\sim 10$ \AA\ boxcar filter.
(b)  The normalized night--sky spectrum over the same observed wavelength
range.  For background--limited observations of faint objects,
night--sky emission lines are the dominant source of noise.}
\label{oned}
\end{figure}

F 36246--1511 was discovered in a 5400s exposure obtained on UT 19
February 1998.  The source appeared as solo emission line spatially offset
by $\sim 2^{\prime\prime}$ from an absorption--line galaxy (F 36247--1510; 
$z = 0.641$).  A portion of the two--dimensional spectrogram, centered on
the emission line, is shown in Figure \ref{serx2}.  The top panel shows
the original two--dimensional spectrogram;  the continuum of the
absorption--line galaxy and the spatially offset emission line can be
readily seen.  In the bottom panel, we have subtracted a Gaussian fit to
the foreground continuum source.  The fit was made to the continuum source
only blueward of the emission line so that after subtraction --- assuming
a locally flat spectrum for both sources --- any remaining flux could be
attributed to the high--redshift candidate.  In this fashion we hoped to
isolate continuum flux from the high--redshift source and recover a
continuum break, which would lend credence to the
Ly$\alpha$--interpretation.  However, as can be seen in the
one--dimensional extracted spectrum (Figure \ref{oned}), the continuum
break is of low significance relative to the noise. 

As the emission line itself is not obviously asymmetric, the remaining
evidence in favor of the Ly$\alpha$--interpretation is two--fold.  To
begin, the observed frame equivalent width of the line is $\sim 300$ \AA. 
This value exceeds the largest equivalent widths observed for other likely
interpretations:  200 \AA\ for the H$\alpha+$[\ion{N}{2}] complex; 100
\AA\ for \oiiib; and 100 \AA\ for \oii\ \citep{stern99}.  Additionally, a
faint source is visible in the Outer West $I_{814}$ Flanking Field image
(W96)  located at the correct separation and orientation to be the
progenitor of the solo emission line.  Unfortunately, as the offset
between the foreground continuum source and the high--redshift candidate
is only $\sim 2^{\prime\prime}$, ground--based images are insufficient to
resolve the two objects.  Hence, the only available visual identification
of the high--redshift candidate stems from the well--resolved but
comparatively shallow single--orbit Flanking Field image. 

Since the discovery spectrum was obtained, we have targeted F 36246--1511
for an additional $\sim 25$ ks of spectroscopy.  The resulting composite
spectrum confirms the $z = 5.631$ interpretation and will appear in a
future work. 

\section{Conclusion}

In the course of our on--going program to study distant galaxies in the
HDF, we have produced as a fringe benefit a deep, serendipitous slit
spectroscopy survey sensitive to a wide range of redshifts.  Our catalog
contains 74 serendipitously detected galaxies, 13 of which are galaxies
in the central HDF which had no prior published spectroscopic redshift, 30
of which are galaxies located in the HDF Flanking Fields.  Five of the
serendipitously detected galaxies are members of a galaxy cluster at $z =
0.85$, and an additional five are candidate Ly$\alpha$--emitters at $z
\simgtr 5$.  The serendipitous sample demonstrates the redshift clustering
behavior observed in other high--redshift samples.  Moreover, our
spectroscopic catalog indicates that photometric redshift techniques
generally compare favorably with spectroscopic redshift determinations. 
As all of the spectra presented herein were obtained entirely without cost
to the main observing campaign, the contribution made by this catalog to
the rich database of observations of the HDF may be regarded as a
testament to the persistent utility of serendipity in observational
astronomy.


\acknowledgments

We are indebted to the expert staff of the Keck Observatory for their
assistance in obtaining the data herein.  It is a pleasure to thank T.
Bida, W. Wack, J. Aycock R. Quick, T. Stickel, G. Punawai, R. Goodrich, R.
Campbell, and B.  Schaeffer for their invaluable assistance during Keck
runs.  We thank N. Brandt and B. Holden for many useful discussions
concerning X--rays.  We are grateful to M. Dickinson for acting as the
steward of published redshifts in the HDF and for providing us with
photometrically--selected targets; to J. Cohen for supporting LRIS and for
making her own redshift survey available; to C. Steidel and A.  Barger for
releasing their optical and IR images of the HDF and its Flanking Fields;
to A. Fern\'{a}ndez--Soto for releasing and maintaining the interactive
HDF catalog of photometric redshifts;  and again to C. Steidel for
supporting LRIS--B.  SD is humbly indebted to JDS, GMR, MEE, JLW, and
especially to EEBG, without whom this work would not have been possible.
The work of DS was carried out at the Jet Propulsion Laboratory,
California Institute of Technology, under contract with NASA.  AJB was   
supported by a NICMOS postdoctoral fellowship while at Berkeley (NASA  
Grant NAG\,5--3043), and a U.K.\ PPARC observational rolling grant
postdoctoral position at the Institute of Astronomy in Cambridge
(ref.~no.~PPA/G/O/1997/00793). HS gratefully acknowledges NSF grant AST
95--28536 for supporting much of the research presented herein.  AD
acknowledges partial support from NASA HF--01089.01--97A and from NOAO.
NOAO is operated by AURA, Inc., under cooperative agreement with the NSF.
This work made use of NASA's Astrophysics Data System Abstract Service.


\eject

{}


\eject


\eject

\begin{deluxetable}{llllll}
\tablewidth{0pt}
\tablecaption{Spectral Categories}
\tablehead{
\colhead{Quality Class} & \colhead{Class Description}}
\startdata
1 & Multiple features \\
2 & Solo line with continuum; assume \oii\ \\
3 & Solo line with continuum break; assume Ly$\alpha$ \\
4 & Solo line with no continuum; assess imaging, if available \\
5 & Continuum break; assess continuum strength blueward of break \\
\enddata
\label{qc}
\end{deluxetable}


\eject

\begin{deluxetable}{lllllcll}
\tablewidth{0pt}
\tablecaption{Serendipitously Detected Galaxies in the Central HDF}
\tablehead{
\colhead{ID\tablenotemark{a}} &
\colhead{$I_{814}$\tablenotemark{a}} &
\colhead{$\alpha_{\mbox{\tiny J}2000}$\tablenotemark{b}} &
\colhead{$\delta_{\mbox{\tiny J}2000}$\tablenotemark{c}} &
\colhead{$z$} &
\colhead{SC\tablenotemark{d}} &
\colhead{References\tablenotemark{e}} &
\colhead{Comments\tablenotemark{f}}}
\startdata
1--95.0   & 24.07 & 36\min45\secper855 & 13\arcmin25\farcs81 & 0.847 & 2 & \nodata  & [\ion{O}{2}] \\
2--201.0  & 23.74 & 36\min47\secper178 & 13\arcmin41\farcs82 & 1.313 & 1 & \nodata  & [\ion{O}{2}], \ion{Mg}{2} abs \\
2--173.0\tablenotemark{\dag}  & 23.45 & 36\min48\secper474 & 13\arcmin16\farcs62 & 0.474 & 1 & \nodata & [\ion{O}{2}], \ion{Ca}{2} H,K abs \\
2--600.0\tablenotemark{\ddag} & 25.59 & 36\min49\secper804 & 14\arcmin19\farcs15 & 0.425 & 4 & \nodata  & [\ion{O}{2}] \\
2--982.0  & 22.70 & 36\min55\secper528 & 13\arcmin53\farcs48 & 1.144 & 1 & C96, P97 & [\ion{O}{2}], \ion{Mg}{2} abs \\
3--318.0  & 24.45 & 36\min54\secper805 & 12\arcmin58\farcs05 & 0.851 & 2 & \nodata  & [\ion{O}{2}] \\
3--342.0  & 24.57 & 36\min58\secper190 & 13\arcmin06\farcs58 & 0.475 & 1 & \nodata  & [\ion{O}{3}]a,b, H$\beta$ \\
3--430.1  & 24.30 & 36\min56\secper603 & 12\arcmin52\farcs70 & 1.233 & 2 & C00      & [\ion{O}{2}] \\
3--773.0  & 22.46 & 36\min57\secper214 & 12\arcmin25\farcs83 & 0.563 & 1 & C96      & [\ion{O}{2}], [\ion{O}{3}]a,b \\
3--863.0  & 23.39 & 36\min58\secper649 & 12\arcmin21\farcs72 & 0.682 & 1 & C96      & [\ion{O}{2}], [\ion{O}{3}]a,b \\
4--131.0  & 24.91 & 36\min49\secper365 & 12\arcmin14\farcs64 & 0.934 & 2 & \nodata  & [\ion{O}{2}] \\
4--236.0\tablenotemark{*} & 28.26 & 36\min47\secper838 & 12\arcmin18\farcs30 & 0.102 & 4 & \nodata  & [\ion{O}{3}]b \\
4--402.1  & 24.96 & 36\min43\secper822 & 12\arcmin51\farcs96 & 1.013 & 2 & \nodata  & [\ion{O}{2}] \\
4--402.3  & 21.13 & 36\min43\secper964 & 12\arcmin50\farcs13 & 0.557 & 2 & C96, C00 & [\ion{O}{2}] \\
4--416.0  & 24.38 & 36\min46\secper555 & 12\arcmin03\farcs09 & 0.454 & 1 & C00      & [\ion{O}{2}], H$\beta$ \\
4--430.0  & 23.30 & 36\min44\secper181 & 12\arcmin40\farcs39 & 0.873 & 4 & C96, C00 & [\ion{O}{2}] \\
4--471.0  & 21.93 & 36\min46\secper511 & 11\arcmin51\farcs32 & 0.503 & 4 & C96      & [\ion{O}{2}] \\
4--491.0\tablenotemark{+} & 24.86 & 36\min43\secper253 & 12\arcmin38\farcs86 & 2.442 & 3 & \nodata  & Ly$\alpha$ \\
4--493.0  & 21.74 & 36\min43\secper156 & 12\arcmin42\farcs20 & 0.849 & 1 & C96, C00 & \ion{Ca}{2} H,K abs, D4000, G band \\
4--565.0  & 22.68 & 36\min43\secper627 & 12\arcmin18\farcs25 & 0.749 & 1 & C96, C00 & [\ion{O}{2}], [\ion{O}{3}]b \\
4--639.1  & 24.65 & 36\min41\secper712 & 12\arcmin38\farcs75 & 2.592 & 3 & S96, C00 & Ly$\alpha$ \\
4--658.0  & 24.77 & 36\min44\secper734 & 11\arcmin43\farcs77 & 0.558 & 1 & \nodata  & [\ion{O}{2}], [\ion{O}{3}]a,b \\
4--727.0  & 23.00 & 36\min43\secper409 & 11\arcmin51\farcs57 & 1.238 & 2 & C00      & [\ion{O}{2}] \\
4--937.0  & 25.09 & 36\min42\secper284 & 11\arcmin26\farcs18 & 0.559 & 1 & \nodata  & [\ion{O}{2}], [\ion{Ne}{3}] \\
4--948\tablenotemark{\times} & 24.99 & 36\min41\secper427 & 11\arcmin42\farcs89 & 1.524 & 2 & \nodata  & [\ion{O}{2}] \\
\enddata
\label{hdfz}
\tablenotetext{a}{Object IDs and $I_{814}$ magnitudes are from \citet{williams96}.}
\tablenotetext{b}{Add 12 hours to the right ascension.}
\tablenotetext{c}{Add 62 degrees to the declination.}
\tablenotetext{d}{See \S \ref{redshifts} and Table \ref{qc}.}
\tablenotetext{e}{
{\it References} lists spectroscopic redshifts already in the literature.
The following abbreviations are used:
C96 $=$ \citet{cohen96}, C00 $=$ \citet{cohen00}, P97 $=$ \citet{phillips97}, S96 $=$ \citet{steidel96a}.}
\tablenotetext{f}{The oxygen emission lines are abbreviated: [\ion{O}{2}] = \oii; [\ion{O}{3}]a = \oiiia; [\ion{O}{3}]b = \oiiib.}
\tablenotetext{\dag}{Listed without a redshift as H36485\_1317 in \citet{cohen00}. Redshift identification tentative; weak detection.}
\tablenotetext{\ddag}{Redshift identification tentative.  Weak detection consistent with \oii--interpretation of solo line;
possible detection of very faint additional lines is roughly consistent with \oiiib--interpretation.}
\tablenotetext{*}{Redshift identification tentative.  Weak detection.
Object colors (see W96) are not consistent with Ly$\alpha$--interpretation of solo
line; \oii--interpretation suggests presence of \oiiib\ at $\lambda_{\mbox{\tiny obs}} = 7415$ \AA, which is not detected;
\oiiib--interpretation suggests presence of H$\alpha$ at $\lambda_{\mbox{\tiny obs}} = 7232$ \AA, which may be very weakly
detected.}
\tablenotetext{+}{Listed as NICMOS \#850 with $\zp = 2.40$ in \citet{dickinson01}.}
\tablenotetext{\times}{The data given are for 4--948.1111, a daughter object likely to be a part of the system formed by 4--948.2,
4--948.11, 4--948.111, 4--948.112, 4--948.1112, 4--948.11111, and 4--948.11112.
This system is distinct from that formed by 4--948.0, 4--948.1, and 4--948.12, which
has a redshift of $z=0.585$ \citep{phillips97,cohen00}.}
\end{deluxetable}


\eject

\renewcommand{\arraystretch}{0.9}

\begin{deluxetable}{lllllcll}
\tablewidth{0pt}
\tablecaption{Serendipitously Detected Galaxies Outside of the Central HDF}
\tablehead{
\colhead{ID} &
\colhead{$I_{\mbox{\tiny AB}}$\tablenotemark{a}} &
\colhead{$\alpha_{\mbox{\tiny J}2000}$\tablenotemark{b}} &
\colhead{$\delta_{\mbox{\tiny J}2000}$\tablenotemark{c}} &
\colhead{$z$} &
\colhead{SC\tablenotemark{d}} &
\colhead{FF\tablenotemark{e}} &
\colhead{Comments\tablenotemark{f}}}
\startdata
\fsz F 36179--1635  &  \fsz     20.1  &  \fsz 36\min17\secper97  &  \fsz 16\arcmin35\farcs0  &  \fsz 0.681  &  \fsz 1  &  \fsz \nodata  &  \fsz [\ion{O}{2}], [\ion{O}{3}]a,b, \ion{Ca}{2} H,K abs \\
\fsz F 36184--1601  &  \fsz     22.3  &  \fsz 36\min18\secper43  &  \fsz 16\arcmin01\farcs6  &  \fsz 0.797  &  \fsz 1  &  \fsz \nodata  &  \fsz [\ion{O}{2}], [\ion{O}{3}]a,b, H$\beta$ \\
\fsz F 36191--6217  &  \fsz $>$ 25.0  &  \fsz 36\min19\secper12  &  \fsz 17\arcmin04\farcs2  &  \fsz 3.896  &  \fsz 4  &  \fsz \nodata  &  \fsz Ly$\alpha$; pstn. from spectrum \\
\fsz F 36197--1601  &  \fsz     22.9  &  \fsz 36\min19\secper78  &  \fsz 16\arcmin01\farcs3  &  \fsz 1.345  &  \fsz 1  &  \fsz \nodata  &  \fsz [\ion{O}{2}], \ion{Mg}{2} abs \\
\fsz F 36218--1513  &  \fsz $>$ 25.0  &  \fsz 36\min21\secper87  &  \fsz 15\arcmin13\farcs7  &  \fsz 5.767  &  \fsz 4  &  \fsz OW  &  \fsz Ly$\alpha$; pstn. from spectrum \\
\fsz F 36219--1516  &  \fsz     24.4  &  \fsz 36\min21\secper91  &  \fsz 15\arcmin16\farcs8  &  \fsz 4.890  &  \fsz 3  &  \fsz OW  &  \fsz Ly$\alpha$ \\
\fsz F 36220--1459  &  \fsz     22.9  &  \fsz 36\min22\secper04  &  \fsz 14\arcmin59\farcs7  &  \fsz 0.849  &  \fsz 2  &  \fsz OW  &  \fsz [\ion{O}{2}] \\
\fsz F 36240--1516  &  \fsz     23.3  &  \fsz 36\min24\secper05  &  \fsz 15\arcmin16\farcs2  &  \fsz 0.796  &  \fsz 2  &  \fsz OW  &  \fsz [\ion{O}{2}] \\
\fsz F 36241--1514  &  \fsz     22.7  &  \fsz 36\min24\secper18  &  \fsz 15\arcmin14\farcs5  &  \fsz 0.222  &  \fsz 1  &  \fsz OW  &  \fsz H$\alpha$, [\ion{O}{3}]b, H$\beta$ \\
\fsz F 36246--1511  &  \fsz $>$ 25.0  &  \fsz 36\min24\secper61  &  \fsz 15\arcmin11\farcs9  &  \fsz 5.631  &  \fsz 4  &  \fsz OW  &  \fsz Ly$\alpha$; pstn. from spectrum \\
\fsz F 36247--1510  &  \fsz     20.1  &  \fsz 36\min24\secper70  &  \fsz 15\arcmin10\farcs5  &  \fsz 0.641  &  \fsz 1  &  \fsz OW  &  \fsz \ion{Ca}{2} H,K, H$\delta$ abs, D4000 \\
\fsz F 36249--1834\tablenotemark{\dag}  &  \fsz \nodata  &  \fsz 36\min24\secper92  &  \fsz 18\arcmin34\farcs1  &  \fsz 0.852  &  \fsz 2  &  \fsz \nodata  &  \fsz [\ion{O}{2}] \\
\fsz F 36255--1510  &  \fsz     22.7  &  \fsz 36\min25\secper50  &  \fsz 15\arcmin10\farcs7  &  \fsz 0.680  &  \fsz 2  &  \fsz OW  &  \fsz [\ion{O}{2}] \\
\fsz F 36265--1443  &  \fsz     24.2  &  \fsz 36\min26\secper58  &  \fsz 14\arcmin43\farcs9  &  \fsz 0.625  &  \fsz 1  &  \fsz OW  &  \fsz [\ion{O}{2}], [\ion{O}{3}]a,b, H$\beta$, H$\gamma$ \\
\fsz F 36270--1509  &  \fsz     20.7  &  \fsz 36\min27\secper04  &  \fsz 15\arcmin09\farcs4  &  \fsz 0.794  &  \fsz 1  &  \fsz OW  &  \fsz \ion{Ca}{2} H,K abs\\
\fsz F 36279--1507  &  \fsz     21.4  &  \fsz 36\min27\secper98  &  \fsz 15\arcmin07\farcs8  &  \fsz 0.680  &  \fsz 2  &  \fsz OW  &  \fsz [\ion{O}{2}] \\
\fsz F 36279--1750\tablenotemark{\dag}  &  \fsz \nodata  &  \fsz 36\min27\secper97  &  \fsz 17\arcmin50\farcs4  &  \fsz 4.938  &  \fsz 4  &  \fsz \nodata  &  \fsz Ly$\alpha$; pstn. from spectrum \\
\fsz F 36289--1752\tablenotemark{\dag}  &  \fsz \nodata  &  \fsz 36\min28\secper93  &  \fsz 17\arcmin52\farcs7  &  \fsz 1.592  &  \fsz 2  &  \fsz \nodata  &  \fsz [\ion{O}{2}] \\
\fsz F 36316--1604  &  \fsz     21.1  &  \fsz 36\min31\secper65  &  \fsz 16\arcmin04\farcs1  &  \fsz 0.785  &  \fsz 2  &  \fsz \nodata  &  \fsz [\ion{O}{2}] \\
\fsz F 36339--1604  &  \fsz     22.4  &  \fsz 36\min33\secper97  &  \fsz 16\arcmin04\farcs7  &  \fsz 0.834  &  \fsz 1  &  \fsz \nodata  &  \fsz [\ion{O}{2}], [\ion{O}{3}]a,b \\
\fsz F 36348--1628  &  \fsz     22.1  &  \fsz 36\min34\secper87  &  \fsz 16\arcmin28\farcs4  &  \fsz 0.847  &  \fsz 1  &  \fsz \nodata  &  \fsz [\ion{O}{2}], \ion{Ca}{2} H,K abs \\    
\fsz F 36356--1424\tablenotemark{\ddag}  &  \fsz 23.1  &  \fsz 36\min35\secper59  &  \fsz 14\arcmin24\farcs0  &  \fsz 2.011  &  \fsz 1  &  \fsz IW  &  \fsz See \S \ref{chandra} \\
\fsz F 36361--1656  &  \fsz     20.9  &  \fsz 36\min36\secper16  &  \fsz 16\arcmin56\farcs9  &  \fsz 0.488  &  \fsz 1  &  \fsz \nodata  &  \fsz [\ion{O}{2}], H$\alpha$ \\
\fsz F 36362--1709  &  \fsz     21.8  &  \fsz 36\min36\secper22  &  \fsz 17\arcmin09\farcs3  &  \fsz 0.945  &  \fsz 2  &  \fsz \nodata  &  \fsz [\ion{O}{2}] \\
\fsz F 36367--1604  &  \fsz     22.6  &  \fsz 36\min36\secper77  &  \fsz 16\arcmin04\farcs8  &  \fsz 0.851  &  \fsz 5  &  \fsz \nodata  &  \fsz D4000 \\                          
\fsz F 36376--1047  &  \fsz     22.3  &  \fsz 36\min37\secper64  &  \fsz 11\arcmin47\farcs8  &  \fsz 0.880  &  \fsz 2  &  \fsz SW  &  \fsz [\ion{O}{2}] \\
\fsz F 36376--1453  &  \fsz     22.4  &  \fsz 36\min37\secper63  &  \fsz 14\arcmin53\farcs7  &  \fsz 4.886  &  \fsz 4  &  \fsz IW  &  \fsz Ly$\alpha$; visual ID uncertain \\
\fsz F 36382--1053  &  \fsz     23.7  &  \fsz 36\min38\secper20  &  \fsz 10\arcmin53\farcs0  &  \fsz 0.766  &  \fsz 2  &  \fsz SW  &  \fsz [\ion{O}{2}] \\
\fsz F 36382--1605  &  \fsz     21.2  &  \fsz 36\min38\secper22  &  \fsz 16\arcmin05\farcs1  &  \fsz 0.852  &  \fsz 1  &  \fsz \nodata  &  \fsz [\ion{O}{2}], D4000 \\                  
\fsz F 36387--1059  &  \fsz     24.8  &  \fsz 36\min38\secper75  &  \fsz 11\arcmin59\farcs3  &  \fsz 3.956  &  \fsz 4  &  \fsz SW  &  \fsz Ly$\alpha$ \\
\fsz F 36397--1547  &  \fsz     21.0  &  \fsz 36\min39\secper76  &  \fsz 15\arcmin47\farcs9  &  \fsz 0.847  &  \fsz 1  &  \fsz \nodata  &  \fsz \ion{Ca}{2} H,K abs, D4000 \\           
\fsz F 36398--1601  &  \fsz     22.8  &  \fsz 36\min39\secper83  &  \fsz 16\arcmin01\farcs6  &  \fsz 0.843  &  \fsz 5  &  \fsz \nodata  &  \fsz D4000 \\                          
\fsz F 36405--1334  &  \fsz     24.1  &  \fsz 36\min40\secper51  &  \fsz 13\arcmin34\farcs9  &  \fsz 3.826  &  \fsz 4  &  \fsz IW  &  \fsz Ly$\alpha$ \\
\fsz F 36417--1437  &  \fsz     23.4  &  \fsz 36\min41\secper72  &  \fsz 14\arcmin37\farcs7  &  \fsz 0.940  &  \fsz 2  &  \fsz IW  &  \fsz [\ion{O}{2}] \\
\fsz F 36452--1108  &  \fsz     23.3  &  \fsz 36\min45\secper24  &  \fsz 11\arcmin08\farcs8  &  \fsz 0.512  &  \fsz 1  &  \fsz SE  &  \fsz [\ion{O}{2}], H$\beta$, [\ion{O}{3}]a,b \\
\fsz F 36466--1517  &  \fsz     24.9  &  \fsz 36\min46\secper68  &  \fsz 15\arcmin17\farcs2  &  \fsz 0.652  &  \fsz 2  &  \fsz NW  &  \fsz [\ion{O}{2}]; visual ID uncertain \\
\fsz F 36488--1500  &  \fsz $>$ 25.0  &  \fsz 36\min48\secper87  &  \fsz 15\arcmin00\farcs6  &  \fsz 2.924  &  \fsz 4  &  \fsz NW  &  \fsz Ly$\alpha$; pstn. from spectrum  \\
\fsz F 36488--1502\tablenotemark{*}  &  \fsz 24.4  &  \fsz 36\min48\secper87  &  \fsz 15\arcmin02\farcs5  &  \fsz 3.111  &  \fsz 3  &  \fsz NW  &  \fsz Ly$\alpha$ \\
\fsz F 36490--1512  &  \fsz     22.7  &  \fsz 36\min49\secper07  &  \fsz 15\arcmin12\farcs4  &  \fsz 0.458  &  \fsz 1  &  \fsz NW  &  \fsz [\ion{O}{2}], H$\beta$ \\
\fsz F 36490--1620  &  \fsz     21.8  &  \fsz 36\min49\secper08  &  \fsz 16\arcmin20\farcs8  &  \fsz 0.501  &  \fsz 2  &  \fsz NW  &  \fsz [\ion{O}{2}] \\
\fsz F 36492--1645  &  \fsz     23.4  &  \fsz 36\min49\secper25  &  \fsz 16\arcmin45\farcs7  &  \fsz 0.536  &  \fsz 4  &  \fsz NW  &  \fsz [\ion{O}{2}] \\
\fsz F 36568--1353  &  \fsz     25.0  &  \fsz 36\min56\secper88  &  \fsz 13\arcmin53\farcs6  &  \fsz 3.43:  &  \fsz 5  &  \fsz NE  &  \fsz Ly break \\
\fsz F 37043--1335  &  \fsz     22.9  &  \fsz 37\min04\secper35  &  \fsz 13\arcmin35\farcs3  &  \fsz 0.592  &  \fsz 1  &  \fsz IE  &  \fsz \ion{Ca}{2} H,K abs; visual ID uncertain \\
\fsz F 37051--1210  &  \fsz     22.5  &  \fsz 37\min05\secper18  &  \fsz 12\arcmin10\farcs7  &  \fsz 0.387  &  \fsz 1  &  \fsz IE  &  \fsz H$\alpha$, [\ion{O}{3}]a,b, H$\beta$ \\
\fsz F 37069--1208  &  \fsz     23.7  &  \fsz 37\min06\secper98  &  \fsz 12\arcmin08\farcs1  &  \fsz 0.693  &  \fsz 1  &  \fsz IE  &  \fsz [\ion{O}{2}], H$\beta$, [\ion{O}{3}]a,b \\
\fsz F 37098--1400  &  \fsz     24.8  &  \fsz 37\min09\secper80  &  \fsz 14\arcmin00\farcs2  &  \fsz 3.910  &  \fsz 3  &  \fsz \nodata  &  \fsz Ly$\alpha$ \\
\fsz F 37131--1333  &  \fsz     21.9  &  \fsz 37\min13\secper11  &  \fsz 13\arcmin33\farcs8  &  \fsz 0.842  &  \fsz 1  &  \fsz \nodata  &  \fsz [\ion{O}{2}], [\ion{O}{3}]a,b \\
\fsz F 37138--1335  &  \fsz     21.5  &  \fsz 37\min13\secper88  &  \fsz 13\arcmin35\farcs2  &  \fsz 0.776  &  \fsz 2  &  \fsz \nodata  &  \fsz [\ion{O}{2}] \\
\fsz F 37180--1248  &  \fsz     22.4  &  \fsz 37\min18\secper06  &  \fsz 12\arcmin48\farcs2  &  \fsz 0.908  &  \fsz 2  &  \fsz OE  &  \fsz [\ion{O}{2}] \\
\enddata
\label{ffz}
\fsz
\tablenotetext{a}{Isophotal magnitude.}
\tablenotetext{b}{Add 12 hours to the right ascension.}
\tablenotetext{c}{Add 62 degrees to the declination.}
\tablenotetext{d}{See \S \ref{redshifts} and Table \ref{qc}.}
\tablenotetext{e}{Indicated galaxy is located in one of the HD\fsz F Flanking Field observations \citep[see][Table 2]{williams96}:
OW = Outer West; SW = South West; IW = Inner West; SE = South East; NW = North West; NE = North East; IE = Inner East; OW = Outer East.}
\tablenotetext{f}{Oxygen emission lines are abbreviated: [\ion{O}{2}] = \oii; [\ion{O}{3}]a = \oiiia; [\ion{O}{3}]b = \oiiib.}
\tablenotetext{\dag}{Indicated galaxy falls outside of the Hawaii 2.2m $I$--band image of \citet{barger99}; the identification was made
in our own 70 minute $R$--band image obtained with ESI (where possible); $I_{\mbox{\tiny AB}}$ magnitudes are not available.}
\tablenotetext{\ddag}{Optical ID for X--ray source CXOHDFN J123635.6$+$621424 \citep{hornschemeier01}.  See \S \ref{chandra}.}
\tablenotetext{*}{Redshift identification tentative.  See discussion of SC 3 in \S \ref{redshifts}.}
\end{deluxetable}


\eject

\renewcommand{\arraystretch}{1}

\begin{deluxetable}{llllcl}
\tablewidth{0pt}
\tablecaption{Summary of Properties of Spectroscopic Members of ClG 1236$+$6215}
\tablehead{
\colhead{ID\tablenotemark{a}} &
\colhead{$\alpha_{\mbox{\tiny J}2000}$\tablenotemark{b}} &
\colhead{$\delta_{\mbox{\tiny J}2000}$\tablenotemark{c}} &
\colhead{z} &
\colhead{$(V-I)_{\mbox{\tiny AB}}$} &
\colhead{Radius\tablenotemark{d}}}
\startdata
F 36348--1628  &   36\min34\secper87  &   16\arcmin28\farcs4  &   0.847  &   1.9  &    9\farcs8 \\
F 36367--1604  &   36\min36\secper77  &   16\arcmin04\farcs8  &   0.851  &   2.4  &   42\farcs7 \\
F 36382--1605  &   36\min38\secper22  &   16\arcmin05\farcs1  &   0.852  &   2.9  &   10\farcs2 \\
F 36397--1547  &   36\min39\secper76  &   15\arcmin47\farcs9  &   0.847  &   2.6  &   10\farcs5 \\
F 36398--1601  &   36\min39\secper83  &   16\arcmin01\farcs6  &   0.843  &   2.5  &   16\farcs8 \\
C 36392--1623  &   36\min39\secper22  &   16\arcmin23\farcs4  &   0.850  &   2.4  &   28\farcs2 \\
C 36421--1545  &   36\min42\secper16  &   15\arcmin45\farcs2  &   0.857  &   1.8  &   25\farcs8 \\
C 36435--1532  &   36\min43\secper50  &   15\arcmin32\farcs2  &   0.847  &   2.7  &   40\farcs3 \\
\enddata
\label{clg}
\tablenotetext{a}{Entries beginning with {\em F} are galaxies described in this catalogue.  Entries beginning with {\em C} are
described in \citet{cohen00}.}
\tablenotetext{b}{Add 12 hours to the right ascension.}
\tablenotetext{c}{Add 62 degrees to the declination.}
\tablenotetext{d}{{\em Radius} indicates the angular distance of the galaxy from the nominal cluster
center: $\alpha=$12\hr36\min39\secper6 $\delta=+$62\deg15\arcmin54$^{\prime\prime}$ (J2000).}
\end{deluxetable}

\end{document}